\documentclass[11pt]{article}

\usepackage{graphicx}
\usepackage{color}
\usepackage{hyperref}
\usepackage[square,numbers,sort&compress]{natbib}

\newcommand{\hept}{$^4$He$(\vec{e},e^\prime\vec{p}\,)^3$H}
\newcommand{\henh}{$^4$He$(\vec{e},e^\prime\vec{n}\,)^3$He}
\newcommand{\hpn}{$^2$H$(\vec{e},e^\prime\vec{p}\,)n$}

\newcommand{\figwidth}{4.25in}

\begin{document}

\begin{center}
\renewcommand{\baselinestretch}{1.2}\normalsize
{\Large\bf Hadrons in the Nuclear Medium}\\[1.5ex]

S.~Strauch$^1$, S.~Malace$^2$, and M.~Paolone$^1$\\
for the Jefferson Lab Hall A Collaboration\\[1ex]
{\em 
$^1$University of South Carolina, Columbia, SC 29208, USA\\
$^2$Duke University, Durham, NC 27708, USA
}

\vspace*{0.5cm}
\end{center}

\begin{abstract}
  Nucleon properties are modified in the nuclear medium.  To
  understand these modifications and their origin is a central issue
  in nuclear physics.  For example, a wide variety of QCD-based
  models, including quark-meson coupling and chiral-quark soliton
  models, predict that the nuclear constituents change properties with
  increasing density.  These changes are predicted to lead to
  observable changes in the nucleon structure functions and
  electromagnetic form factors.

  We present results from a series of recent experiments at MAMI and
  Jefferson Lab, which measured the proton recoil polarization in the
  \hept{} reaction to test these predictions.  These results, with the
  most precise data at $Q^2 = 0.8$ (GeV/$c$)$^2$ and at 1.3
  (GeV/$c$)$^2$ from E03-104, put strong constraints on available
  model calculations, such that below $Q^2 = 1.3$ (GeV/$c$)$^2$ the
  measured ratios of polarization-transfer are successfully described
  in a fully relativistic calculation when including a medium
  modification of the proton form factors or, alternatively, by strong
  charge-exchange final-state interactions.  We also discuss possible
  extensions of these studies with measurements of the \hept{} and
  \hpn{} reactions as well as with the neutron knockout in \henh{}.
\end{abstract}

\section{Introduction}

The underlying theory of strong interactions is Quantum ChromoDynamics
(QCD), yet there are no ab-initio calculations of nuclei
available. Nuclei are effectively and well described as clusters of
protons and neutrons held together by a strong, long-range force
mediated by meson exchange, whereas the saturation properties of
nuclear matter arise from the short-range, repulsive part of the
strong interaction \cite{Moszkowski:1960zz}. At nuclear densities of
about 0.17 nucleons/fm$^3$, nucleon wave functions have significant
overlap. In the chiral limit, one expects nucleons to lose their
identity altogether and nuclei to make a transition to a quark-gluon
plasma \cite{Pisarski:1981mq}.  This phase transition is extensively
being studied at the RHIC facility.

Within QCD, there is no known way to derive anything like an atomic
nucleus in which the constituents do not change as the mean density
(or temperature) goes away from zero \cite{AWThomasPC03}.  The
discovery of the nuclear EMC effect, the depletion of the deep
inelastic structure function observed in the valence quark regime,
almost twenty years ago brought the subject of quarks into nuclear
physics with great impact.  However, the specific causes of the
modifications observed in the nuclear structure functions have not yet
been identified with certainty \cite{Sargsian:2002wc}.  Miller and
Smith \cite{Miller:2001tg} argue that the depletion is due to some
interesting effect involving dynamics beyond the conventional
nucleon-meson treatment of nuclear physics.  One such explanation is a
medium modification of bound nucleon structure. A variety of models
predict deviations from the free-space nucleon form factors in the
nuclear medium: A calculation by Lu {\it et
  al.}~\cite{Lu:1997mu,Lu:1998tn}, using a quark-meson coupling (QMC)
model, suggests deviations from the free-space
electromagnetic form factor which result in measurable effects on observables in
model calculations over the four-momentum-transfer squared,
$Q^2$, range 0.0 $< Q^2 <$ 2.5 (GeV/$c$)$^2$. Similar measurable
effects have been calculated in a light-front-constituent quark model
by Frank {\it et al.}  \cite{Frank:1995pv}, a modified Skyrme model by
Yakshiev {\it et al.}~\cite{Yakhshiev:2002sr}, a chiral quark-soliton
model (CQS) by Smith and Miller \cite{Smith:2004dn}, and the
Nambu-Jona-Lasinio model \cite{Horikawa:2005dh,Cloet:2009tx}.  These
calculations are generally consistent with present constraints on
possible medium modifications for both the electric form factor (from
the Coulomb Sum Rule, for Q$^2$ $<$ 0.5 (GeV/$c$)$^2$
\cite{Jourdan:1995jz,Morgenstern:2001jt,Carlson:2003}) and the
magnetic form factor (from a $y$-scaling analysis \cite{Sick:1988wu}
for Q$^2$ $>$ 1 (GeV/$c$)$^2$), and limits on the scaling of nucleon
magnetic moments in nuclei \cite{Ericson:1986zz}.

Although models using free nucleons and mesons as quasi-particles are
successful in the description of many aspects of nuclear physics, one
may therefore expect that their use is under certain circumstances a
highly uneconomical approach, especially given that these are not the
fundamental entities of the underlying theory. The
use of medium-modified nucleons as quasi-particles may be a better
choice.  To experimentally demonstrate any modification of the nucleon
form factors, one is required to have excellent control over the
reaction mechanism effects \cite{VanOrden:2006vi}.  The nucleus, as a
bound many-body quantum system, has inherent many-body effects, such
as meson-exchange currents (MEC) and isobar configurations (IC). In
addition, when probing nuclear structure one has to deal with
final-state interactions (FSI). Thus, distinguishing possible changes
in the spatial structure of nucleons embedded in a nucleus from more
conventional many-body effects is only possible within the context of
a model.

\section{Recoil Polarization in Quasi-Elastic Electron Scattering}

\begin{figure}[ht]\centering
\includegraphics[width=\figwidth]{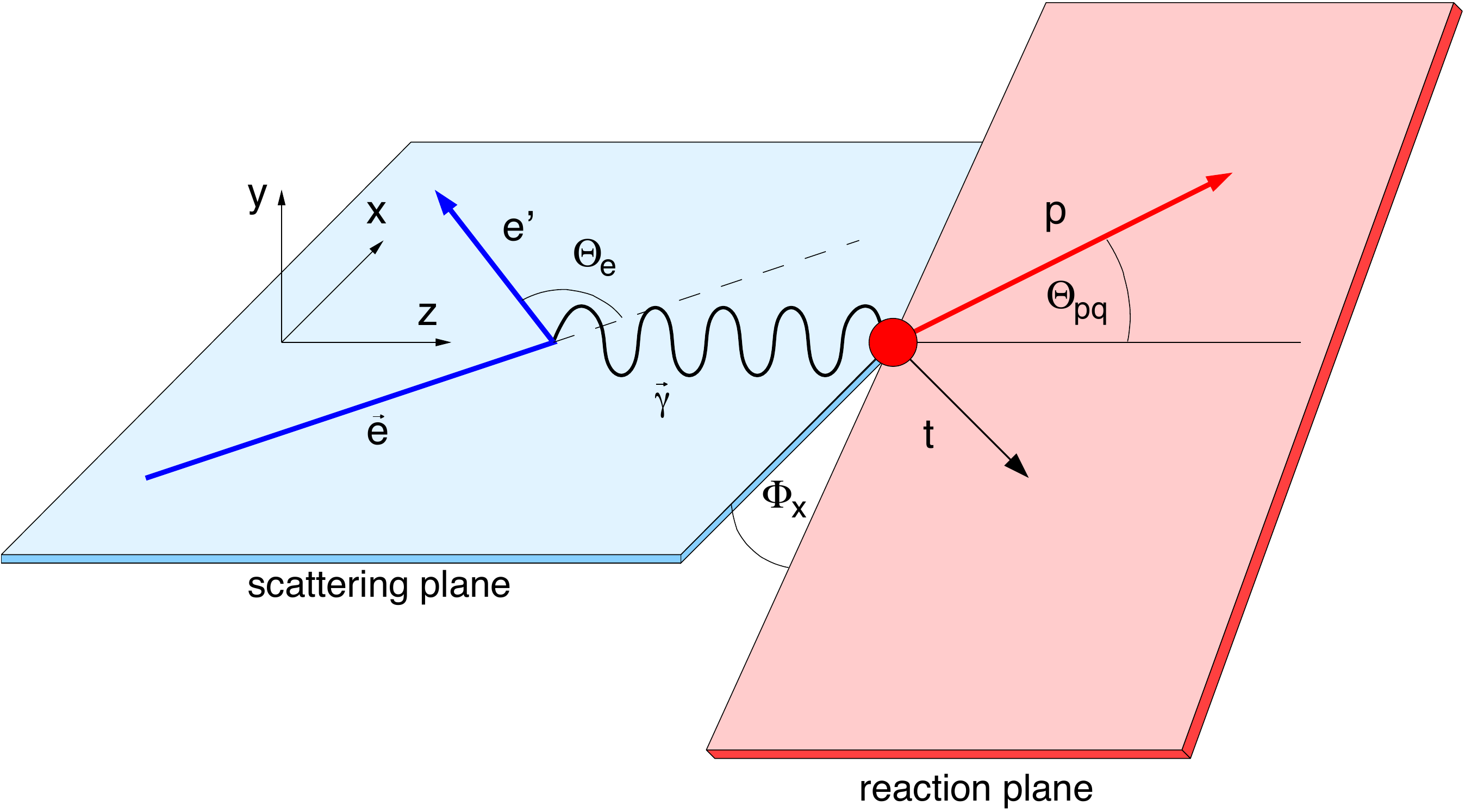}
\caption{Coordinate system used to define the components of the recoil
proton polarization in the \hept{} reaction. The $z$ axis is along the
momentum transfer, the $x$ axis is in the scattering plane
perpendicular to the momentum transfer $\vec q$ and the $y$ axis is
perpendicular to the scattering plane, forming a right-handed system.}
\label{fig:kinem}
\end{figure}

The charge and magnetic responses of a single nucleon are quite well
studied from elastic scattering experiments. Measuring the same
response from quasi-elastic scattering off nuclei and comparing with a
single nucleon is thus an intuitive method to investigate the
properties of nucleons inside nuclei. In free electron-nucleon
scattering, the ratio of the electric to magnetic Sachs form factors,
$G_E$ and $G_M$, is given by \cite{Akhiezer:1974em,Arnold:1980zj}:
\begin{equation}
\frac{G_E}{G_M} = -\frac{P'_x}{P'_z} \cdot \frac{E_e + E_{e'}}{2m_p}
\tan (\theta_e /2),
\label{eq:free}
\end{equation}
where $P'_x$ and $P'_z$ are the transverse and longitudinal
transferred polarizations; see Fig.~\ref{fig:kinem}.  The beam energy
is $E_e$, the energy (angle) of the scattered electron is $E_{e'}$
($\theta_e$) and $m_p$ is the proton mass.  This relation was
extensively used to extract $G_E/G_M$ for the proton, see {\it e.g.}
\cite{Jones:1999rz,Gayou:2001qt,Gayou:2001qd,Ron:2007vr,Puckett:2010rz}
for measurements at JLab.

When such measurements are performed on a nuclear target in
quasi-elastic kinematics, the experimental results for the
polarization-transfer ratio are conveniently expressed in terms of the
polarization double ratio
\begin{equation}
R = \frac{(P_x'/P_z')_{A}}{(P_x'/P_z')_{^1\rm H}},
\label{eq:rexp}
\end{equation}
where the polarization-transfer ratio for the quasi-elastic proton
knockout $A(\vec e, e'\vec p)$ reaction is normalized to the
polarization-transfer ratio measured in elastic $^1$H$(\vec e,e'\vec
p)$ scattering in the identical setting in order to emphasize
differences between the in-medium and free values. Such a
double ratio cancels also nearly all experimental systematic
uncertainties.  A proper interpretation of the results requires
accounting for such effects as FSI and MEC. At high momentum transfer,
however, the contribution of many-body and rescattering mechanisms are
strongly suppressed \cite{Laget:1994xx}.  Polarization-transfer
observables provide us with a way to study the behavior of the nucleon
form factors in the nuclear medium.

\section{Present Experimental Results}

\begin{figure}[tb]
\begin{center}
\includegraphics[width=\figwidth]{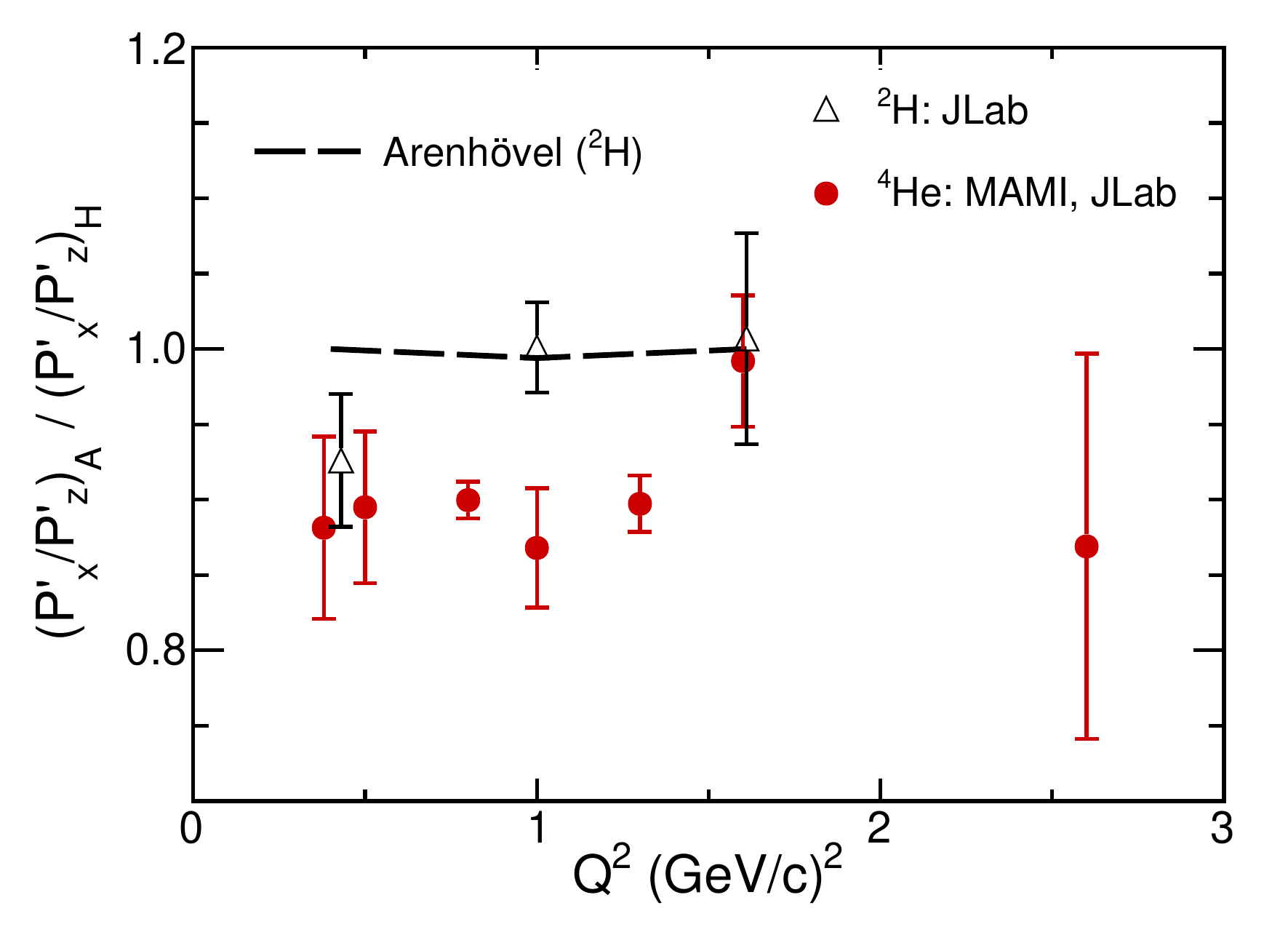}
\caption{
Bound-to-free polarization-transfer double ratio $R$ at low missing
momenta for $^2$H$(\vec e, e' \vec
p)n$ (open triangles) from \protect\cite{Hu:2006fy} and for $^4$He$(\vec e, e'
\vec p)^3$H (closed circles) from
\protect\cite{Dieterich:2000mu,Strauch:2002wu,Paolone:2010qc} as a function of
$Q^2$. The curve shows a result of a calculation by Arenh\"ovel (dashed line)
for deuterium.
\label{fig:ratioplotA}}
\end{center}
\end{figure}

JLab Experiment E89-033 was the first to measure the polarization
transfer in a complex nucleus, $^{16}$O \cite{Malov:2000rh}.  The
results are consistent with predictions of relativistic calculations
based on the free-proton form factor with an experimental uncertainty
of about 18\%.  Polarization transfer has been used previously to
study nuclear medium effects in deuterium
\cite{Eyl:1995fk,Milbrath:1997de,Barkhuff:1999xc}.  Within statistical
uncertainties, no evidence of medium modifications was found. More
recently, polarization-transfer data on $^2$H were measured in JLab
experiment E89-028 \cite{Hu:2006fy}, under conditions very similar to
those for experiment E93-049 \cite{Strauch:2002wu} on
$^4$He. Realistic calculations to describe this reaction were
performed by Arenh\"ovel. Experimental results (open triangles) for
the $^2$H-to-$^1$H polarization-transfer double ratio, along with the
results of a calculation by Arenh\"ovel (dashed curve), are shown in
Fig.~\ref{fig:ratioplotA}.
Arenh\"ovel's full calculation describes the $^2$H data well.  As the
sampled density is small and the bound proton in $^2$H is nearly on
mass-shell, it is not surprising that there are no indications for
medium modifications of the proton electromagnetic form factors in the
$^2$H data.

One might expect to find larger medium effects in $^4$He, with its
significantly higher density. Indeed, recent Jefferson Lab Experiment
E03-103 has measured the EMC effect for various nuclei and results
indicate that the nuclear dependence of the cross section is nearly
identical for $^4$He and $^{12}$C \cite{Seely:2009gt}.  Although
estimates of the many-body effects in $^4$He may be more difficult
than in $^2$H, calculations for $^4$He indicate they are small
\cite{Laget:1994xx}.  The first $^4$He$(\vec e,e^\prime \vec p)^3$H
proton recoil-polarization measurements were performed at MAMI at $Q^2
= 0.4$ (GeV/$c$)$^2$ \cite{Dieterich:2000mu} and at Jefferson Lab Hall
A at $Q^2$ = 0.5, 1.0, 1.6, and 2.6 (GeV/$c$)$^2$, E93-049
\cite{Strauch:2002wu}. Experiment E03-104 added two high-precision
points at $Q^2$ = 0.8 and 1.3 (GeV/$c$)$^2$ \cite{Paolone:2010qc}.
The results are shown in Fig.~\ref{fig:ratioplotA} (solid points).
The missing-mass technique was used to identify $^3$H in the final
state.  For a reliable interpretation of the experimental data it is
imperative to have good control over conventional many-body effects in
the reaction.  All these data were thus taken in quasi-elastic
kinematics at low missing momentum with symmetry about the
three-momentum-transfer direction to minimize these effects.
Furthermore, they can be studied with the induced polarization, $P_y$,
which is a direct measure of final-state
interactions. Induced-polarization data were taken simultaneously to
the polarization-transfer data.

\begin{figure}[h!tb]
\begin{center}
  \includegraphics[width=\figwidth]{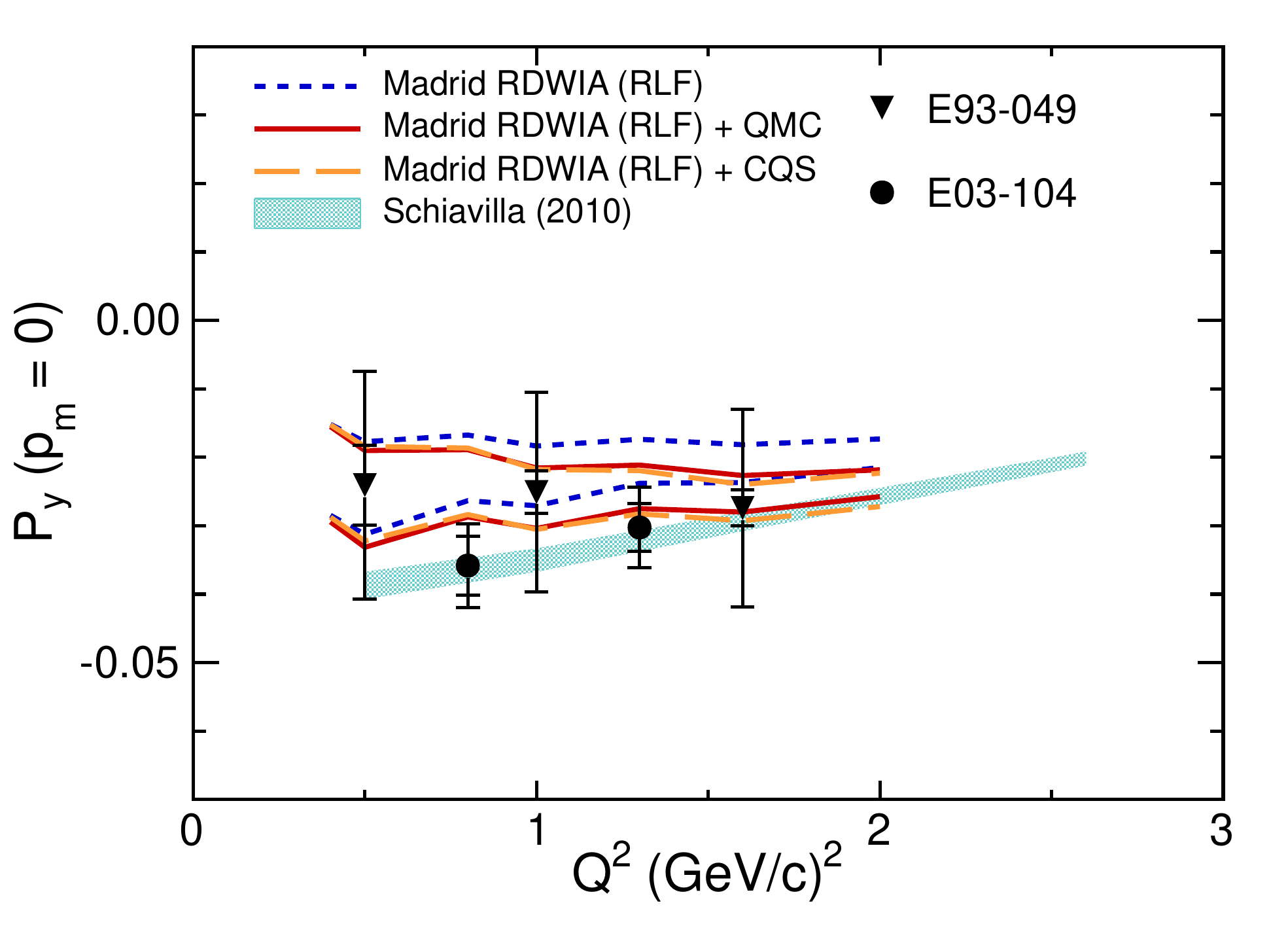}
  \caption{$^4$He$(e,e^\prime \vec p)^3$H induced polarization data
    from Jefferson Lab experiment E93-049 \cite{Strauch:2002wu} along with
    results from experiment E03-104 \cite{Malace:2010}. The data are compared
    to calculations from Schiavilla {\it et al.}  \cite{Schiavilla:2004xa}
    and the Madrid group \cite{Udias:1999tm,Caballero:1997gc,Udias:2000ig}
    using the {\it cc1} (lower set of curves) and {\it cc2} (upper set of
    curves) current operators.  In-medium form factors from the QMC
    \cite{Lu:1997mu} (solid curve) and CQS \cite{Smith:2004dn} (dashed curve)
    models were used in two of the Madrid calculations.  Note that the
    comparison is made for missing momentum $p_m \approx 0$ and that the
    experimental data have been corrected for the spectrometer acceptance for
    this comparison.  }
  \label{fig:py}
\end{center}
\end{figure}

Figure \ref{fig:py} shows the results for $P_y$. The induced
polarization is small at the low missing momenta in this measurement.
\begin{figure}[h!tb]
\begin{center}
  \includegraphics[width=\figwidth]{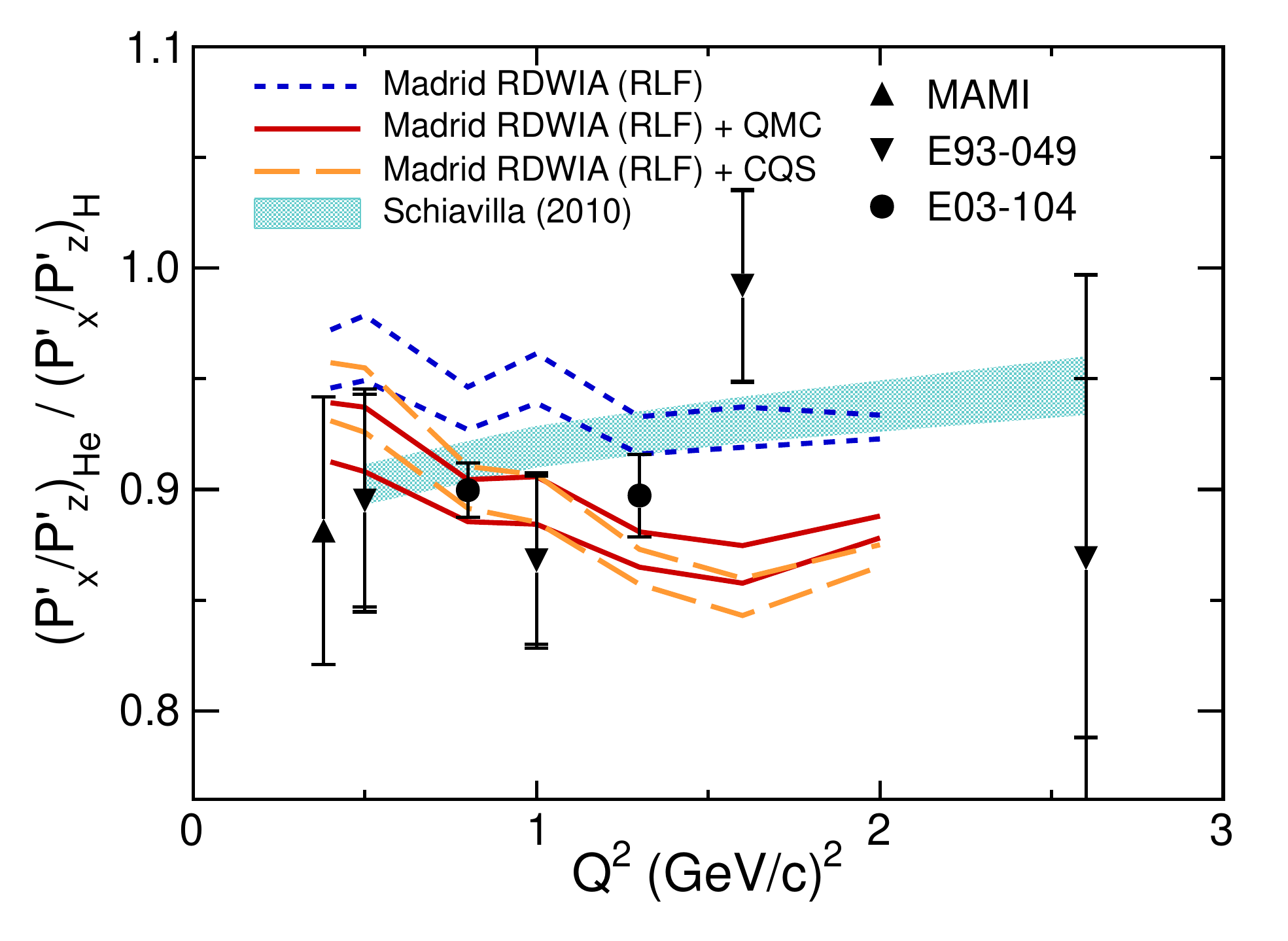}
  \caption{$^4$He$(\vec e, e'\vec p)^3$H
    polarization-transfer double ratio $R$ as a function of
    $Q^2$ from Mainz \cite{Dieterich:2000mu} and Jefferson Lab experiments
    E93-049 \cite{Strauch:2002wu} (open symbols) 
    and E03-104 \cite{Paolone:2010qc} (filled circles).  The data are
    compared to calculations from the Madrid group
    \cite{Udias:1999tm,Caballero:1997gc,Udias:2000ig}, using the {\it cc1}
    (lower set of curves) and {\it cc2} (upper set of curves) current
    operators, and
    Schiavilla {\it et al.} \cite{Schiavilla:2004xa} as in Fig.~\ref{fig:py}.
    Not shown are a relativistic Glauber model calculation by the Ghent group
    \cite{Lava:2004mp} and results from Laget \cite{Laget:1994xx} which give 
    both a value of $R\approx 1$.
    \label{fig:ratio}}
\end{center}
\end{figure}
The sizable systematic uncertainties are due to possible instrumental
asymmetries.  Dedicated data were taken during E03-104 to study these
and help significantly reduce systematic uncertainties in the
extraction of $P_y$. The data are compared with results of a
relativistic distorted-wave impulse approximation (RDWIA) calculation
by the Madrid group
\cite{Udias:1999tm,Caballero:1997gc,Udias:2000ig}. In this model FSI
are incorporated using an updated version of the RLF relativistic
optical potentials \cite{Horowitz:1985tw,Murdock:1986fs} that distort
the final nucleon wave function; the MRW optical potential of
\cite{McNeil:1983yi}, used in \cite{Paolone:2010qc}, does not yield an
as good description of $P_y$ as the modified RLF potential shown here.
Charge-exchange terms are not taken into account in the Madrid RDWIA
calculation; preliminary studies show, however, that they are of
small effect in this model \cite{UdiasPC10}.  Calculations are shown
for choices of {\it cc1} and {\it cc2} current operators as defined in
\cite{DeForest:1983vc}.  The choice {\it cc1} yields the largest
prediction for $P_y$ in absolute value and describes the data well;
possibly hinting at the importance of the lower spinor components in
this relativistic calculation; see \cite{Udias:2000ig}.  We note that
these RDWIA calculations provide also good descriptions of, e.g., the
induced polarizations as measured at Bates in the $^{12}$C(e,e$^\prime
\vec p$) reaction \cite{Woo:1998zz,Udias:2000ig} and of $A_{TL}$ in
$^{16}$O($e, e^\prime p$) as previously measured at JLab
\cite{Gao:2000ne}.  While the polarization-transfer observables are
expected to be sensitive to possible nucleon medium modifications,
results of the RDWIA calculation including medium-modified form
factors show only some small effect on the induced polarization.  The
data are also compared with the results of a calculation from
Schiavilla {\it et al.}  \cite{Schiavilla:2004xa} (shaded band).  That
model uses variational wave functions for the bound three- and
four-nucleon systems, non-relativistic MEC and free nucleon form
factors. The FSIs are treated within the optical potential framework
and include both spin-independent and spin-dependent charge-exchange
terms which play a crucial role in the prediction of $P_y$.  Note that
the charge-exchange term gives the largest contribution to
Schiavilla's calculation of P$_{y}$.  This model describes the data
well after being constrained to the new data from E03-104.

The $^4$He polarization-transfer double ratio is shown in Figure
\ref{fig:ratio}. The recent data from E03-104 (filled circles)
\cite{Paolone:2010qc} are consistent with the previous data from
E93-049 \cite{Strauch:2002wu} and MAMI \cite{Dieterich:2000mu} (open
symbols). The polarization-transfer ratio $(P'_x/P'_z)$ in the $(\vec
e,e' \vec p)$ reaction on $^4$He is significantly different from those
on hydrogen.  The data are compared with results of the same
RDWIA calculations by the Madrid group
\cite{Udias:1999tm,Caballero:1997gc,Udias:2000ig} (dotted curves) as
in Fig.~\ref{fig:py}.  MEC are not explicitly included in the Madrid
calculation. Predictions by Meucci {\it et al.} \cite{Meucci:2002iy}
show that the two-body current (the seagull diagram) effects on the
polarization-transfer ratio are generally small; less than 3 \% at low
missing momenta and visible only at high missing momenta.  It can be
seen that the Madrid RDWIA calculation (dotted curves) overpredicts
the data.  The agreement of the Madrid model with the
polarization-transfer data is improved after including the
density-dependent medium-modified form factors from the QMC
\cite{Lu:1997mu} or CQS \cite{Smith:2004dn} models in the RDWIA
calculation (solid and dashed curves).  This agreement has been
interpreted as possible evidence of proton medium modifications
\cite{Strauch:2002wu}.  An alternative interpretation of the observed
suppression of the polarization-transfer ratio is offered within the
more traditional calculation by Schiavilla {\it et al.}
\cite{Schiavilla:2004xa} (shaded band). Schiavilla's calculation uses
free nucleon form factors and explicitly includes MEC effects which
are suppressing $R$ by almost 4\%.

Currently, the $^4$He$(\vec e, e^\prime \vec p)^3$H
polarization-transfer data can be well described by either the
inclusion of medium-modified form factors or strong charge-exchange
FSI in the models.  The difference in the modeling of final-state
interactions is the origin of the major part of the difference between
the results of the calculations by Madrid {\it et al.}
\cite{Udias:1999tm,Caballero:1997gc,Udias:2000ig} and Schiavilla {\it
  et al.}~\cite{Schiavilla:2004xa} for the polarization observables.
Optical potentials in these models have now been constrained to the
new induced polarization data from E03-104.  

\section{Possible Future Experiments}

Current quite different, state of the art, models which employ free
nucleon form factors (Madrid RDWIA and Schiavilla) agree in their
predictions above $Q^2 = 1.3$ (GeV/$c$)$^2$ where ambiguities in the
choice of the current operator become smaller.  Including medium
modifications of the proton form factors in the Madrid calculations,
on the other hand, results in an easily observable reduction of the
polarization-transfer double ratio of at least 5\%.  A recent
experiment proposal, PR12-11-002 \cite{PR11002}, to Jefferson Lab PAC
37 therefore proposed to measure one new high-precision data point of
the $^4$He polarization-transfer double ratio at $Q^2 = 1.8$
(GeV/$c$)$^2$.  Such a data point will be decidedly valuable to refute
either of these approaches: If a new result agrees with that set of
calculations without the need of medium modified form factors it will
seriously challenge the present models of in-medium effects.  If, on
the other hand, the new data will agree with those predictions which
include the QMC or CQS form factors it is very hard to see how this
observation can be reconciled in the other models given the already
tight constraints from E03-104.

\begin{figure}[htb!]
\begin{center}
\includegraphics[width=\figwidth]{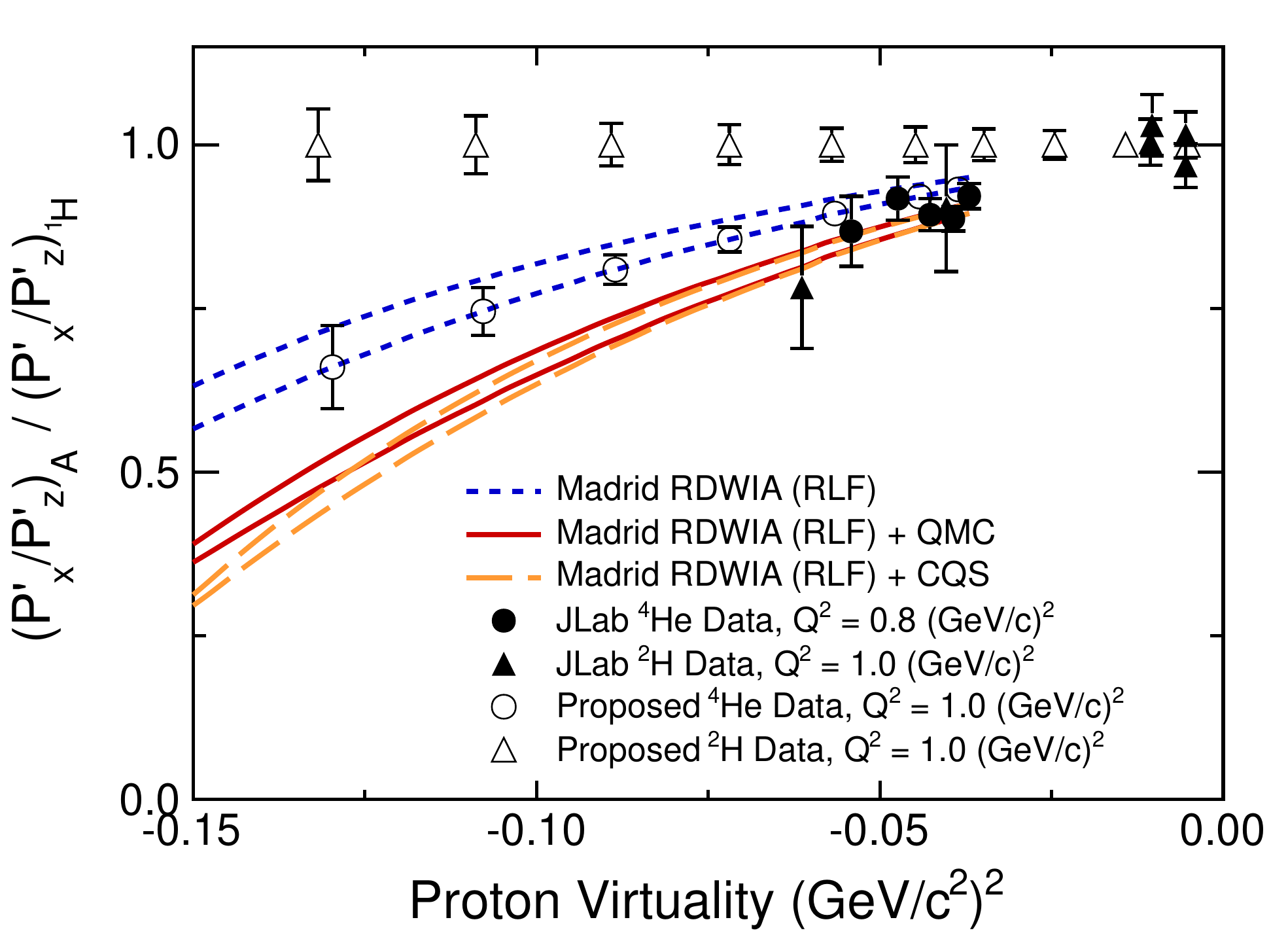}
\caption{\label{fig:q10virt}
 Polarization-transfer ratio $P'_x/P'_z$ from bound nucleon
  knockout off $^4$He and $^2$H 
  compared to $P'_x/P'_z$ from elastic $ep$ scattering as a
  function of proton virtuality.   The curves are
   various calculations using the model of Udias {\it et al.} for the
   reaction on $^4$He and current operators $cc1$ (lower set of curves) and
   $cc2$ (upper set of curves).
  The points indicate previous data \cite{Hu:2006fy,Paolone:2010qc}
  (solid symbols) and the
  statistical uncertainties of the proposed data and which are arbitrarily
  placed on the RDWIA ($cc1$) curve for $^4$He and at $R=1$ for $^2$H.
}
\end{center}
\end{figure}

As second part of PR12-11-002 is an extensive study of the proton
recoil-polarization observables as a function of missing momentum or
proton virtuality in both the \hept{} and \hpn{} reactions at $Q^2 =
1.0$ (GeV/$c$)$^2$.  Ciofi degli Atti {\it et al.}
\cite{Ciofi:2007vx} argue that the modification of the wave function
of the bound nucleon in a nucleus should strongly depend on the
momentum of the nucleon.  The data on deuterium will provide a link
between free $ep$ scattering and quasi-elastic proton knockout in
$^4$He. The $^2$H and $^4$He data have in common that in both cases
the reaction takes place on a bound, off-shell nucleus.  While the
proton in helium is tightly bound in a nuclear medium which is much
denser than that in deuterium similar proton virtualities can be
reached in both, the $^2$H$(e,e'p)n$ and $^4$He$(e,e'p)^3$H reactions
at larger missing momenta; $\approx 300$~MeV/$c$ in the proposed
experiment.  Here, the proton virtuality is defined as $v = p^2 -
m_p^2$, where $p$ is the four-momentum of the bound proton.  In the
impulse approximation $p^2 = (m_A - E_m)^2 - p_m^2$, where $E_m$ and
$p_m$ are respectively the missing energy and momentum in the
$A(e,e'p)$ reaction.  The origin of medium effects, as density
dependent or bound-nucleon-momentum dependent, could thus be studied
in the comparison between both of these data.  Figure
\ref{fig:q10virt} shows $R$ for previous and for the proposed data as
a function of the proton virtuality.

A complementary and very important experiment would be the measurement
of the nucleon knockout in quasielastic scattering in the \henh{}
reaction. Clo\"{e}t {\em et al.}  \cite{Cloet:2009tx} have studied
possible in-medium changes of the bound neutron electromagnetic
form-factor ratio with respect to the free ratio, the superratio
$\left(G^*_E/G^*_M\right)/\left(G_E/G_M\right)$.  At small values of
$Q^2$ this superratio depends on the in-medium modifications of the
nucleon magnetic moment and the effective electric and magnetic radii.
The superratio of the neutron is dominated by the expected increase of
the electric charge radius in the nuclear medium and is found to be
greater than one.  In contrast, the proton superratio is predicted to
be smaller than one.  A comparison of high-precision measurements of
the reactions $^2$H$(\vec e, e' \vec n)p$ and \henh{} would allow to
test these predictions.

\section{Summary}

Polarization transfer in the quasi-elastic proton-knockout reaction is
arguably one of the most direct experimental methods to identify
nuclear-medium changes to nucleon properties, which are predicted by
QCD-based models, as other conventional medium effects, such as
many-body currents and final state interactions, are
suppressed. Furthermore, the possible role of FSI in the
interpretation of these data can be constrained by the induced
polarization $P_y$.  After such constraints, present $^4$He$(\vec e,
e^\prime \vec p)^3$H polarization-transfer data can be well described
by either the inclusion of medium-modified form factors or strong
charge-exchange FSI in the models.  

Possible future measurements of the quasielastic $(\vec e, e'\vec p)$
and $(\vec e, e'\vec n)$ reactions off both $^4$He and $^2$H targets
would allow to further probe the bound nucleon electromagnetic
current, including possible medium modifications of the proton
electromagnetic form factor.

\section{Acknowledgments}

This work was supported in parts by the U.S. National Science Foundation: NSF PHY-0856010.
Jefferson Science Associates operates the Thomas Jefferson National Accelerator Facility under DOE
contract DE-AC05-06OR23177.

\bibliographystyle{mystyle}
\bibliography{references}

\begin{thebibliography}{10}
\providecommand{\url}[1]{\texttt{#1}}
\providecommand{\urlprefix}{URL }
\providecommand{\eprint}[2][]{\url{#2}}

\bibitem{Moszkowski:1960zz}
S.~A. Moszkowski and B.~L. Scott, Annals Phys. \textbf{11} (1960) 65.

\bibitem{Pisarski:1981mq}
R.~D. Pisarski, Phys. Lett. \textbf{B110} (1982) 155.

\bibitem{AWThomasPC03}
A.W.~Thomas, private communication (2003).

\bibitem{Sargsian:2002wc}
M.~M. Sargsian \emph{et~al.}, J. Phys. \textbf{G29} (2003) R1,
  \eprint{nucl-th/0210025}.

\bibitem{Miller:2001tg}
G.~A. Miller and J.~R. Smith, Phys. Rev. \textbf{C65} (2002) 015211,
  \eprint{nucl-th/0107026}.

\bibitem{Lu:1997mu}
D.-H. Lu, A.~W. Thomas, K.~Tsushima, A.~G. Williams, and K.~Saito, Phys. Lett.
  \textbf{B417} (1998) 217, \eprint{nucl-th/9706043}.

\bibitem{Lu:1998tn}
D.-H. Lu, K.~Tsushima, A.~W. Thomas, A.~G. Williams, and K.~Saito, Phys. Rev.
  \textbf{C60} (1999) 068201, \eprint{nucl-th/9807074}.

\bibitem{Frank:1995pv}
M.~R. Frank, B.~K. Jennings, and G.~A. Miller, Phys. Rev. \textbf{C54} (1996)
  920, \eprint{nucl-th/9509030}.

\bibitem{Yakhshiev:2002sr}
U.~T. Yakhshiev, U.-G. Meissner, and A.~Wirzba, Eur. Phys. J. \textbf{A16}
  (2003) 569, \eprint{nucl-th/0211055}.

\bibitem{Smith:2004dn}
J.~R. Smith and G.~A. Miller, Phys. Rev. \textbf{C70} (2004) 065205,
  \eprint{nucl-th/0407093}.

\bibitem{Horikawa:2005dh}
T.~Horikawa and W.~Bentz, Nucl. Phys. \textbf{A762} (2005) 102,
  \eprint{nucl-th/0506021}.

\bibitem{Cloet:2009tx}
I.~C. Cloet, G.~A. Miller, E.~Piasetzky, and G.~Ron, Phys. Rev. Lett.
  \textbf{103} (2009) 082301, \eprint{0903.1312}.

\bibitem{Jourdan:1995jz}
J.~Jourdan, Phys. Lett. \textbf{B353} (1995) 189.

\bibitem{Morgenstern:2001jt}
J.~Morgenstern and Z.~E. Meziani, Phys. Lett. \textbf{B515} (2001) 269,
  \eprint{nucl-ex/0105016}.

\bibitem{Carlson:2003}
J.~Carlson, J.~Jourdan, R.~Schiavilla, and I.~Sick, Phys. Lett. \textbf{B553}
  (2003) 191.

\bibitem{Sick:1988wu}
I.~Sick, Comments Nucl. Part. Phys. \textbf{A18} (1988) 109.

\bibitem{Ericson:1986zz}
T.~E.~O. Ericson and A.~Richter, Phys. Lett. \textbf{B183} (1987) 249.

\bibitem{VanOrden:2006vi}
J.~W. Van~Orden, Phys. Rev. \textbf{C74} (2006) 034607,
  \eprint{nucl-th/0605031}.

\bibitem{Akhiezer:1974em}
A.~I. Akhiezer and M.~P. Rekalo, Sov. J. Part. Nucl. \textbf{4} (1974) 277.

\bibitem{Arnold:1980zj}
R.~G. Arnold, C.~E. Carlson, and F.~Gross, Phys. Rev. \textbf{C23} (1981) 363.

\bibitem{Jones:1999rz}
M.~K. Jones \emph{et~al.}, Phys. Rev. Lett. \textbf{84} (2000) 1398,
  \eprint{nucl-ex/9910005}.

\bibitem{Gayou:2001qt}
O.~Gayou \emph{et~al.}, Phys. Rev. \textbf{C64} (2001) 038202.

\bibitem{Gayou:2001qd}
O.~Gayou \emph{et~al.}, Phys. Rev. Lett. \textbf{88} (2002) 092301,
  \eprint{nucl-ex/0111010}.

\bibitem{Ron:2007vr}
G.~Ron \emph{et~al.}, Phys. Rev. Lett. \textbf{99} (2007) 202002,
  \eprint{0706.0128}.

\bibitem{Puckett:2010rz}
A.~J.~R. Puckett \emph{et~al.}, Phys. Rev. Lett. \textbf{104} (2010) 242301,
  \eprint{1005.3419}.

\bibitem{Laget:1994xx}
J.-M. Laget, Nucl. Phys. \textbf{A579} (1994) 333.

\bibitem{Malov:2000rh}
S.~Malov \emph{et~al.}, Phys. Rev. \textbf{C62} (2000) 057302.

\bibitem{Eyl:1995fk}
D.~Eyl \emph{et~al.}, Z. Phys. \textbf{A352} (1995) 211.

\bibitem{Milbrath:1997de}
B.~D. Milbrath \emph{et~al.}, Phys. Rev. Lett. \textbf{80} (1998) 452,
  \eprint{nucl-ex/9712006}.

\bibitem{Barkhuff:1999xc}
D.~H. Barkhuff \emph{et~al.}, Phys. Lett. \textbf{B470} (1999) 39.

\bibitem{Hu:2006fy}
B.~Hu \emph{et~al.}, Phys. Rev. \textbf{C73} (2006) 064004,
  \eprint{nucl-ex/0601025}.

\bibitem{Strauch:2002wu}
S.~Strauch \emph{et~al.}, Phys. Rev. Lett. \textbf{91} (2003) 052301,
  \eprint{nucl-ex/0211022}.

\bibitem{Dieterich:2000mu}
S.~Dieterich \emph{et~al.}, Phys. Lett. \textbf{B500} (2001) 47,
  \eprint{nucl-ex/0011008}.

\bibitem{Paolone:2010qc}
M.~Paolone \emph{et~al.}, Phys. Rev. Lett. \textbf{105} (2010) 072001,
  \eprint{1002.2188}.

\bibitem{Seely:2009gt}
J.~Seely \emph{et~al.}, Phys. Rev. Lett. \textbf{103} (2009) 202301,
  \eprint{0904.4448}.

\bibitem{Malace:2010}
S.~P. Malace, M.~Paolone, S.~Strauch, \emph{et~al.}, {Submitted to Phys. Rev.
  Lett.} \eprint{1011.4483}.

\bibitem{Schiavilla:2004xa}
R.~Schiavilla, O.~Benhar, A.~Kievsky, L.~E. Marcucci, and M.~Viviani, Phys.
  Rev. Lett. \textbf{94} (2005) 072303, \eprint{nucl-th/0412020}.

\bibitem{Udias:1999tm}
J.~M. Udias, J.~A. Caballero, E.~Moya~de Guerra, J.~E. Amaro, and T.~W.
  Donnelly, Phys. Rev. Lett. \textbf{83} (1999) 5451, \eprint{nucl-th/9905030}.

\bibitem{Caballero:1997gc}
J.~A. Caballero, T.~W. Donnelly, E.~Moya~de Guerra, and J.~M. Udias, Nucl.
  Phys. \textbf{A632} (1998) 323, \eprint{nucl-th/9710038}.

\bibitem{Udias:2000ig}
J.~M. Udias and J.~R. Vignote, Phys. Rev. \textbf{C62} (2000) 034302,
  \eprint{nucl-th/0007047}.

\bibitem{Lava:2004mp}
P.~Lava, J.~Ryckebusch, B.~Van~Overmeire, and S.~Strauch, Phys. Rev.
  \textbf{C71} (2005) 014605, \eprint{nucl-th/0407105}.

\bibitem{Horowitz:1985tw}
C.~J. Horowitz, Phys. Rev. \textbf{C31} (1985) 1340.

\bibitem{Murdock:1986fs}
D.~P. Murdock and C.~J. Horowitz, Phys. Rev. \textbf{C35} (1987) 1442.

\bibitem{McNeil:1983yi}
J.~A. McNeil, L.~Ray, and S.~J. Wallace, Phys. Rev. \textbf{C27} (1983) 2123.

\bibitem{UdiasPC10}
J.M.~Udias, private communication (2010).

\bibitem{DeForest:1983vc}
T.~De~Forest, Nucl. Phys. \textbf{A392} (1983) 232.

\bibitem{Woo:1998zz}
R.~J. Woo \emph{et~al.}, Phys. Rev. Lett. \textbf{80} (1998) 456.

\bibitem{Gao:2000ne}
J.~Gao \emph{et~al.}, Phys. Rev. Lett. \textbf{84} (2000) 3265.

\bibitem{Meucci:2002iy}
A.~Meucci, C.~Giusti, and F.~D. Pacati, Phys. Rev. \textbf{C66} (2002) 034610,
  \eprint{nucl-th/0205055}.

\bibitem{PR11002}
Jefferson Lab Proposal PR12-11-002, "Proton Recoil Polarization in the
  $^4$He(e,e'p)$^3$H, $^2$He(e,e'p)$^3$H, and $^4$He(e,e'p)$^3$H Reactions",
  E.~Brash, G.M.~Huber, R.~Ransome, and S.~Strauch, spokespersons.

\bibitem{Ciofi:2007vx}
C.~Ciofi~degli Atti, L.~L. Frankfurt, L.~P. Kaptari, and M.~I. Strikman, Phys.
  Rev. \textbf{C76} (2007) 055206, \eprint{0706.2937}.

\end{thebibliography}

\end{document}